# Single vortex-antivortex pair in an exciton polariton condensate

Georgios Roumpos[1], Michael D. Fraser[2,3], Andreas Löffler[4], Sven Höfling[4], Alfred Forchel[4] & Yoshihisa Yamamoto[1,2]

*[1]E. L. Ginzton Laboratory, Stanford University, Stanford, CA, USA, [2]National Institute of Informatics, Tokyo, Japan, [3]Institute for Nano Quantum Information Electronics, University of Tokyo, Tokyo, Japan, [4]Technische Physik, University of Würzburg Wilhelm-Conrad-Röntgen-Research Center for Complex Material Systems, Germany*

**In a homogeneous two-dimensional system at non-zero temperature, although there can be no ordering of infinite range[1,2], a superfluid phase is predicted for a Bose liquid[3-5]. The stabilization of phase in this superfluid regime is achieved by the formation of bound vortex-antivortex pairs. It is believed that several different systems share this common behaviour, when the parameter describing their ordered state has two degrees of freedom[6], and the theory has been tested for some of them[7-12]. However, there has been no direct experimental observation of the phase stabilization mechanism by a bound pair. Here we present an experimental technique that can identify a single vortex-antivortex pair in a two-dimensional exciton polariton condensate. The pair is generated by the inhomogeneous pumping spot profile, and is revealed in the time-integrated phase maps acquired using Michelson interferometry, which show that the condensate phase is only locally disturbed. Numerical modelling based on open dissipative Gross-Pitaevskii equation suggests that the pair evolution is quite different in this non-equilibrium system compared to atomic condensates[13]. Our results demonstrate that the exciton polariton condensate is a unique system for studying two-dimensional superfluidity[14-19] in a previously inaccessible regime.**

Microcavity exciton polaritons[20] behave as a system of strictly two-dimensional bosons when their density is below the exciton saturation density. Because of their half-light half-matter nature, their effective mass is extremely small, so that quantum many-body effects are important at relatively high temperatures, even up to room temperature[21]. In particular, dynamic polariton condensation is observed[22-24], and its signatures are similar to Bose-Einstein condensate (BEC), namely massive occupation of the ground state and long range phase coherence. However, the short lifetime allows only the formation of quasi-equilibrium steady state, in which polaritons escaping from the cavity are continuously replenished by the external pump.

Here, we show that single vortex-antivortex pairs can be observed in an exciton polariton condensate. We have created a pumping spot which generates a minimum of the condensate density at the centre. A zero in density can be thought of as a superposition of a vortex and antivortex that can then be separated by an external perturbation[25,26]. Thus, the centre of the condensate acts as a source of vortex-antivortex pairs. We use a Michelson interferometer to reconstruct the time-integrated phase map of the system. When the sample disorder potential is strong, pinned pairs appear at certain locations. When the sample disorder potential is weak, on the other hand, they are mobile. They appear along a fixed axis, because of a small asymmetry in the pumping spot, and are created with a random polarization, namely the relative positions of vortex and antivortex can swap from shot to shot. In the time-integrated measurement, two distinct characteristic phase defects appear in both cases. We show direct experimental evidence that the phase is indeed stabilized away from the core of the pair.

We have applied a dissipative model similar to that in Ref. 16, and we found that vortex-antivortex pair motion is significantly modified because of the steady-state dissipative nature of the polariton condensate and the repulsive interactions between condensate particles and the reservoir or thermal excitations. The vortex pair is found to migrate perpendicular to its dipole moment and in our experimental parameter space, recombines before reaching the condensate edge. Despite the short polariton lifetime, a vortex pair survives for a long enough time to be observed.

Our experiment is performed in a GaAs-based microcavity sample[17,27]. The pump laser at above bandgap energy creates free electon-hole pairs, which form excitons and finally relax towards the lower polariton (LP) branch. When a LP decays from the cavity as a photon, energy and in-plane momentum conservation are satisfied. A schematic of our Michelson interferometer setup is shown in Fig. 1a. We can overlap the real-space image with its reflected version on the camera, and because the two phase fronts are tilted with respect to each other, an interference pattern is observed (Fig. 1b). Changing the difference in the interferometer arm lengths shifts the fringes on the camera. Therefore, the intensity measured at one particular pixel shows a sinusoidal modulation (Fig. 1c). This reveals two pieces of information: the relative phase of the two overlapping images at this pixel point as well as the fringe visibility, which corresponds to the first order spatial correlation function. Repeating this procedure for every pixel point, we can construct both phase and fringe visibility maps.

We employ a commercial beam shaper to create a flat-top pumping spot (see Fig. 5 in the Supplementary Information). Below threshold, luminescence shows Airy-like patterns because of diffraction effects in the laser pumping system. Above the condensation threshold, a population dip develops at the condensate centre. Our numerical simulations suggest that the inhomogeneity of the pump spot profile, notably the ring near the spatial origin, generates this condensate shape. The condensate minimum is stabilized by the repulsive reservoir-condensate interaction, since the reservoir shows a corresponding density maximum at the same point (see Fig. 14 in the Supplementary Information). Phase fluctuations[28] are locally maximized at the condensate centre, where the condensate density is minimum and the non-condensate density (reservoir polaritons) is maximum. Thus, the probability for vortex-antivortex pair excitation is also maximum at this point. Given that the condensate size, and in particular the central density dip responsible for vortex pair formation, is not much larger than the size of the vortex pair it is not possible for more than one vortex pair to enter into the system.

The expected phase map for a pinned vortex-antivortex pair is shown in Fig. 2a. The phase is strongly perturbed around the pair, but is stabilized away from the pair, which is an essential mechanism of the Berezinskii-Kosterlitz-Thouless (BKT) phase ordering. If we perform an interferometry measurement, in the way described above, with this phase map as an input, the expected result is Fig. 2b. The two double-dislocation patterns, one at the upper part, and the other at the lower part of the figure, mark the position of the pair in the original and in the reflected images. Fig. 2c shows a phase map measured experimentally. It features the double dislocation pattern characteristic of a pinned vortex-antivortex pair. In Fig. 2d, we have subtracted the global phase slope to reveal the actual phase map of a single vortex-antivortex pair, corresponding to the expected one (Fig. 2a).

In this case (Fig. 2c-d), the pair has a pinned position and polarization, due to the sample disorder potential. This is similar to the experiment in Ref. 16, where single vortices pinned by the disorder potential were observed. However, the disorder potential in our sample is very weak (standard deviation $\Delta V_{dis} = 71\mu eV$ for potential fluctuations with characteristic length longer than $1\mu m$, see Fig. 10 in the Supplementary Information), and there are only a few spots where pinned vortex-antivortex pairs can be observed. If we mount the sample carefully, so as to minimize strain (see Methods), we can no longer find any pinned vortex pair.

Fig. 3 illustrates typical experimental data after removing the sample strain, which can be qualitatively understood by the simple model described below. We consider a pair along the horizontal direction and at the centre of the condensate (along the dashed line in Fig. 2a). We also assume that pairs are formed with random polarization, namely the vortex and antivortex can swap their relative positions. Based on this model, we simulate the time-integrated experimental result by ensemble averaging. The theoretical phase and fringe visibility maps generated by this simulation are shown in Fig. 3a and b respectively. There are two areas where the phase is shifted by π, one at the top and the other at the bottom of the figure, and they are surrounded by minima in the fringe visibility.

We indeed observed such unique patterns experimentally when the sample strain was removed, as shown in Fig. 3c,d. The shape of the π-phase-shift areas is different from the expected one, because the pair is not pinned at the central position but undergoes a micro-motion. Our experimental data integrate over a distribution for the pair position, and also the pair size. In Fig. 3e, we have not subtracted the global phase slope, so that the direct experimental phase map is given. In Fig. 3f, we plot the cross section of the phase maps along two lines passing through the two areas of π-phase shift, where this phase shift is shown quantitatively. We also plot the measured fringe visibility along the same lines, and find that the visibility minima coincide with the phase jumps. This result demonstrates that the vortex-antivortex pairs are created with a random polarization after the removal of strain. Moreover, when we rotate the prism by $90^o$, so that the reflected image is along the vertical axis, no phase defect is observed (Fig. 3g). The difference between Fig. 3e and Fig. 3g can be understood as follows: The vortex pair always remains on the horizontal axis and we fold the reflected image along the vertical axis. In this case, the phase rotation around the vortex by 2π and that around the antivortex by -2 π cancel out, thus there is no phase defect in the interference pattern (See Fig. 13 in the Supplementary Information). The corresponding visibility map is shown in Fig. 3h.

We consistently observe pairs along the same axis, even after we rotate the sample by 90°. However, when we rotate the pump laser spot by 90°, the polarization axis of the vortex pair is also rotated by 90° as shown in Fig. 9 in the Supplementary Information. These results suggest that the pair polarization direction is determined by a small asymmetry of the pump laser spot, rather than by the disorder potential landscape in the sample. Finally, we have created a gaussian pumping spot and did not observe any phase defects (Fig. 7 of Supplementary Information). This observation supports our argument that the minimum of the condensate density at the centre of the spot acts as a source of vortex-antivortex pairs.

Using a time-dependent open-dissipative Gross-Pitaevskii equation[29], we can numerically study the evolution of a vortex pair imprinted along the horizontal axis. In a harmonically trapped conservative condensate, a single vortex pair will undergo linear motion at a velocity inversely proportional to its separation and upon interaction with the boundary will wrap back upon itself with a cyclical motion (Fig. 4a). However, a polariton condensate with considerable condensate-reservoir interaction (through stimulated scattering and repulsive interactions) experiences drag forces strongly perturbing the vortex pair motion[30], causing it to recombine after only a short travel distance (Fig. 4b). For the types of micro-motion depicted in Figs. 4a and 4b, the interference fringes are still preserved in time-integrated measurements despite the vortex pair motion. The measured positions of the phase defects simply represent the time averaged positions of the vortex and antivortex. The time-integrated phase and visibility maps for the case of a polariton condensate are given in Figs. 4c and 4d respectively.

In conclusion, we created and observed a single vortex-antivortex pair in a two-dimensional exciton polariton condensate. Pairs are generated at the density minimum in the centre of the condensate, and are evidenced in the time-integrated phase maps acquired using Michelson interferometry, which show a localized phase disturbance and good agreement with theoretical predictions. The pair does not rotate, but is created along a fixed axis. It moves along the vertical direction until it recombines to disappear. Our experimental observations are reproduced by the numerical simulation using the open dissipative Gross-Piaevskii equation with gain and loss terms.

**Methods summary**

We employ a commercial beam shaper to create a flat-top pumping spot. The laser pump is vertically polarized and is focused on the sample through a polarizing cube beam splitter (PBS) (see Supplementary Information). Luminescence of the orthogonal linear polarization is detected and split into two parts by a 50-50 non-polarizing cube beam splitter (NPBS). We use a dielectric mirror (M1) for the first arm of the interferometer, and an uncoated glass right-angle prism (M2) for the second one. The position of the prism is controlled by a combination of a translation stage and a piezoelectric actuator. Our imaging system combines a high numerical aperture (NA) objective lens (NA=0.55) with a plano-convex lens and offers a magnification of 50. The image is focused on a high resolution electron-multiplying charge-coupled-device (CCD) camera with pixel size of $8x8\mu m^2$.

**Acknowledgements** We thank M. Kasevich and A. L. Fetter for stimulating discussions. This work was supported by Special Coordination Funds for Promoting Science and Technology, DARPA Navy/SPAWAR N66001-09-1-2024, NICT and MEXT.

**Author Contributions** G.R. designed and built the experimental setup, took and analyzed the data, and wrote the paper. M.D.F. performed the numerical calculations and wrote the paper. A.L., S.H. and A.F. fabricated the microcavity sample. Y.Y. Conceived the experiment. All authors contributed toward data interpretation.

**Author Information** Correspondence and requests for materials should be addressed to G.R. (roumpos@stanford.edu) or Y.Y. (yyamamoto@stanford.edu).

## Methods

**Sample.** Our microcavity sample features a λ/2 AlAs cavity with 3 stacks of 4 GaAs quantum wells at the central three antinodes of the resonant electric field. It shows a Rabi splitting of $2\hbar\Omega_R = 14meV$ . Measurements are performed for photon-exciton detuning $\delta{\sim}0meV$ , which gives an effective mass for lower polaritons (LP) of $m^*_{LP} \sim 5x10^{-5}m_e$. From the linewidth of the reflection dip as well as from measurement of time-resolved luminescence, we estimate the LP lifetime to be $\tau_{LP} \sim 4ps$. The sample is mounted using silver paint on a copper sample holder, attached to the cold finger of a helium flow cryostat. The energy splitting between orthogonally polarized luminescence at $0^o$ collection angle (ground state splitting $\Delta E_{gr}$) can be up to $150\mu eV$ . However, when special care is taken to reduce strain during cooldown, by applying a homogeneous layer of silver paint between the sample and sample holder, the ground state splitting can be reduced below 50μeV[25]. Data in Fig. 2 are taken with $\Delta E_{gr} = 150\mu eV$ , whereas all other data, including supplementary information, are taken with $40\mu eV \leq \Delta E_{gr} \leq 50\mu eV$ . In the latter case, luminescence above threshold is weakly linearly polarized along the direction of minimum energy, with a degree of linear polarization ≤ 20%.

**Laser.** We pump the sample with a modelocked Ti:Sapph laser operated in the continuous wave (CW) mode. To avoid sample heating, we use an optical chopper to modulate the laser in 0.5ms pulses at 100Hz repetition rate. The laser is incident from the normal direction and its wavelength is at the first reflectivity minimum of the microcavity above the stop band (723*nm*). The incident power is controlled by a polarization filtering setup employing a variable retarder, as in Ref. 27. The threshold power of 20*mW* (see Supplementary Information) gives a threshold particle density of ~ $10^3$μm$^{-2}$ for our 20μm-diameter pumping spot, given the 4ps LP lifetime.

**Real space and momentum space imaging.** Real- and momentum-space images are collected through the same setup as in Ref. 31, with a high-NA objective lens (NA=0.55), and CCD camera with pixel size 9.5 x 9.5μm$^2$. For real space, the system offers a magnification of 25, and the diffraction limit is 061 λ/NA = 0.85μm. The momentum-space resolution is 0.019μm$^{-1}$. For real space images, luminescence is filtered through a combination of two interference filters one long-pass at 750*nm* and one bandpass at 770 ± 5*nm*, which block the laser wavelength without distorting the signal spectrum. These experimental results are shown in the Supplementary Information.

FIG. 1: **Michelson Interferometer. a**, Schematic of the setup for measurement of phase and fringe visibility maps. It employs a mirror (M1) and a right-angle prism (M2), which creates the reflection of the original image along one axis, depending on the prism orientation. A two-lens microscope setup overlaps the two real space images of the polariton condensate on the camera. **b**, Typical interference pattern observed above the polariton condensation threshold (at 60mW) along with a schematic showing the orientation of the two overlapping

images. **c**, Blue circles: Intensity on one pixel of the camera as a function of the prism (M2) position in normalized units. Red line: Fitting to a sine function.

FIG. 2: **Phase map of a pinned pair. a**, Calculated phase map of a condensate including a single vortex-antivortex pair. Arrows show the direction of the phase increase around the vortex and antivortex. **b**, Simulation of the experimentally measured phase map when a interferes with its reflection along the horizontal (dashed) line. A global phase slope along the vertical direction is added. **c**, Experimentally measured phase map at 55mW above the condensation threshold of 20mW. The blue square marks the position of a double dislocation pattern. **d**, Expanded view of the blue square in c, where the global slope along the horizontal direction is subtracted. c and d are rotated by 90° with respect to all other experimental data.

FIG. 3: **Phase and fringe visibility map of a free pair. a**, Expected phase map for a pair of fixed size and along the horizontal axis pinned at the centre, but switching positions of the vortex and antivortex. **b**, Corresponding fringe visibility map. **c**, and **d**, Measured phase and fringe visibility maps. **e**, Same as in c, but now the global phase slope is not subtracted. **f**, Blue: phase cross section along the continuous and dashed lines in c. Red: fringe visibility cross section along the same lines. **g**, Measured phase map when the prism is rotated by 90°, along with a schematic showing the orientation of the interfering images. **h**, Corresponding fringe visibility map. Experimental data are taken at 55mW, above the condensation threshold of 20mW.

FIG. 4: **Single vortex pair dynamics. a** and **b** show a sequence of frames depicting the time evolution of a single vortex pair in a harmonically-trapped conservative condensate and a dissipative condensate in the parameter space corresponding to our experiment, the latter pumped by an experimentally

measured laser profile. The time after vortex pair imprinting is indicated. Figures **c** and **d** show the numerical time-integrated fringe visibility and phase maps respectively for the dissipative condensate case (compare with experiments in Figs. 3d,e).

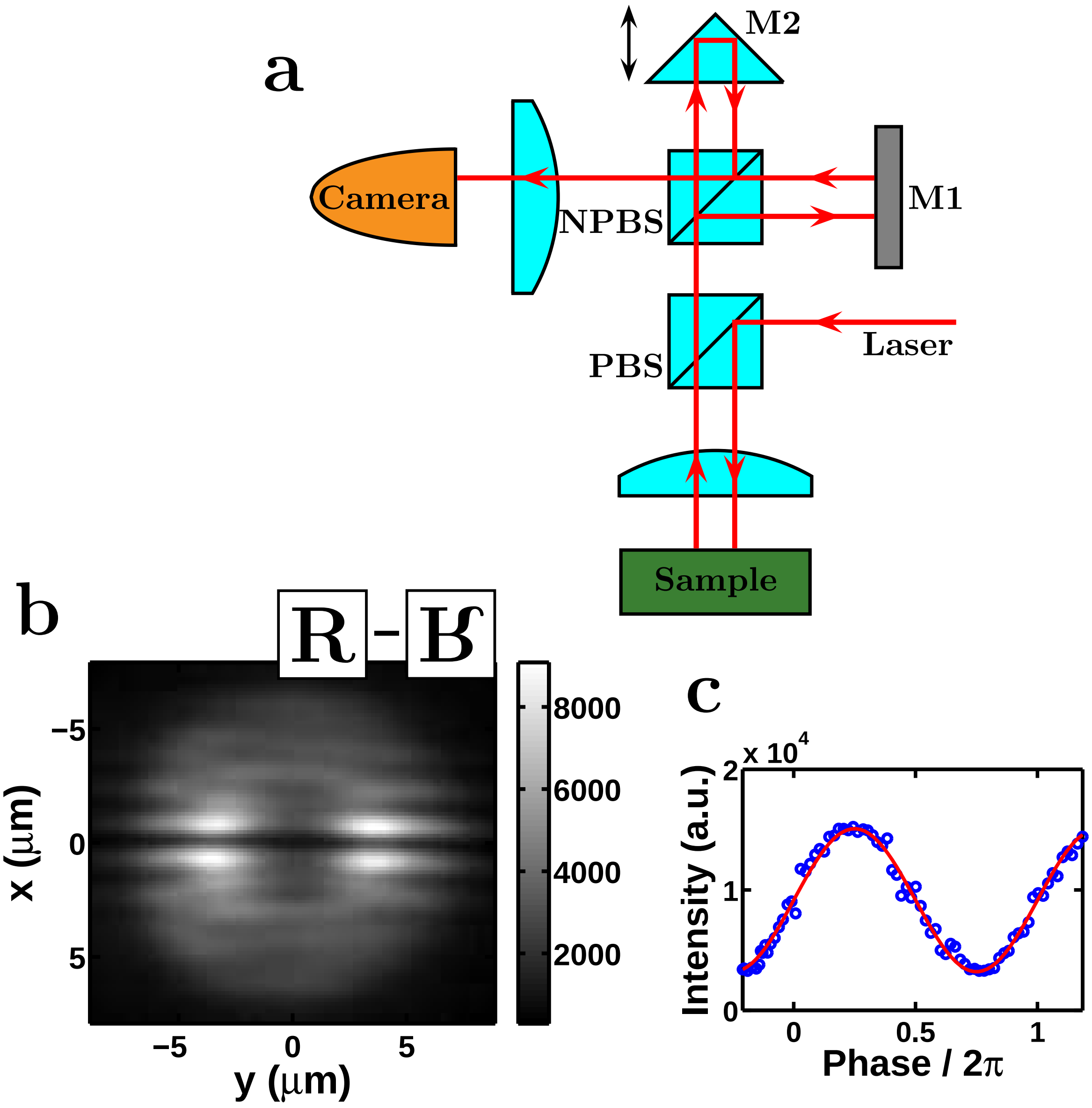

Figure 1:

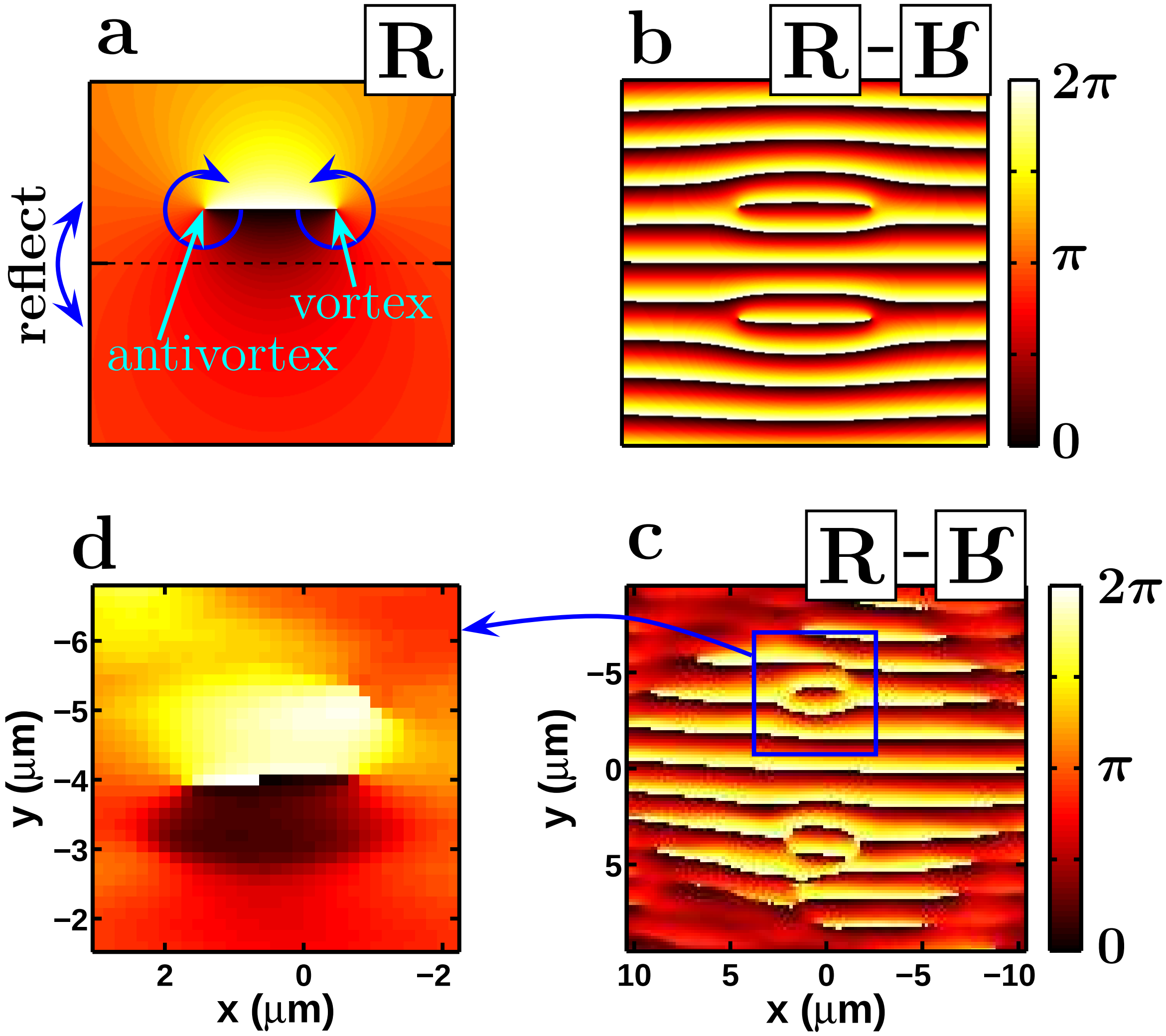

Figure 2:

Figure 3:

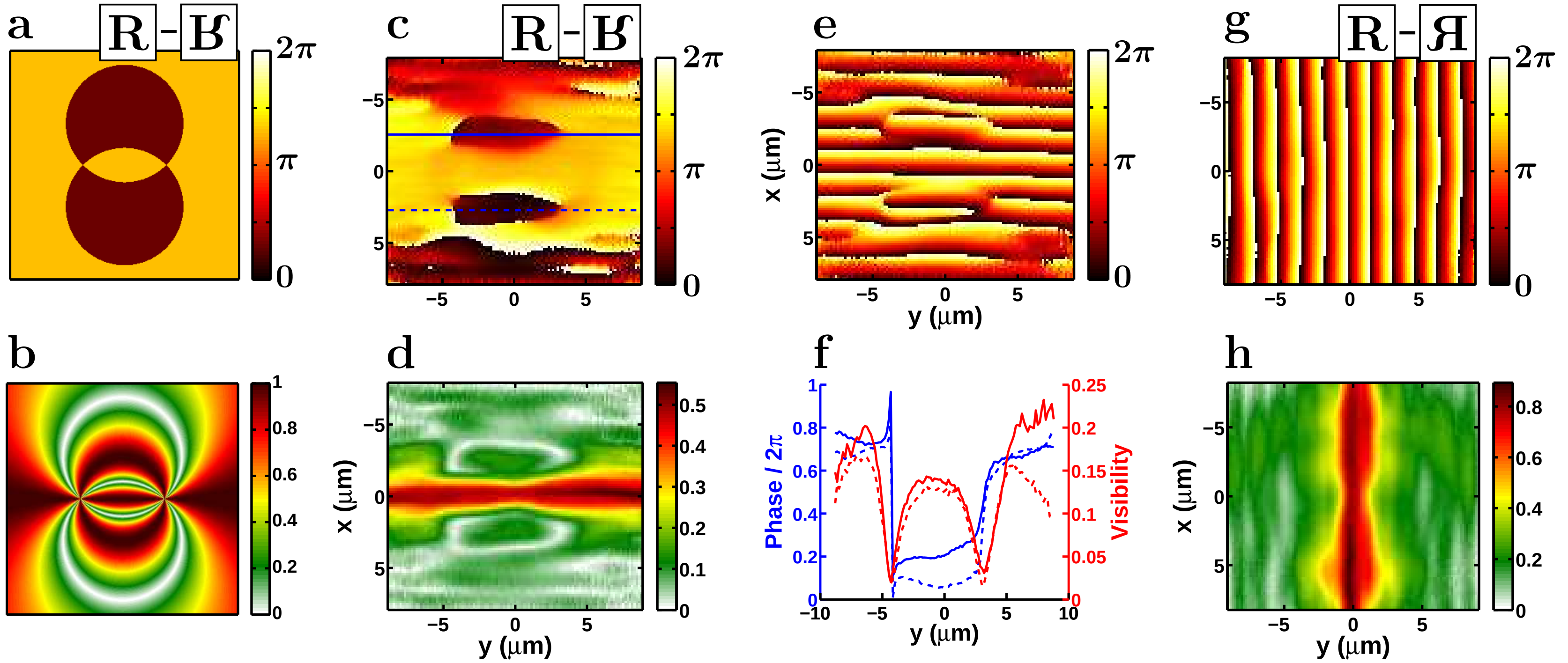

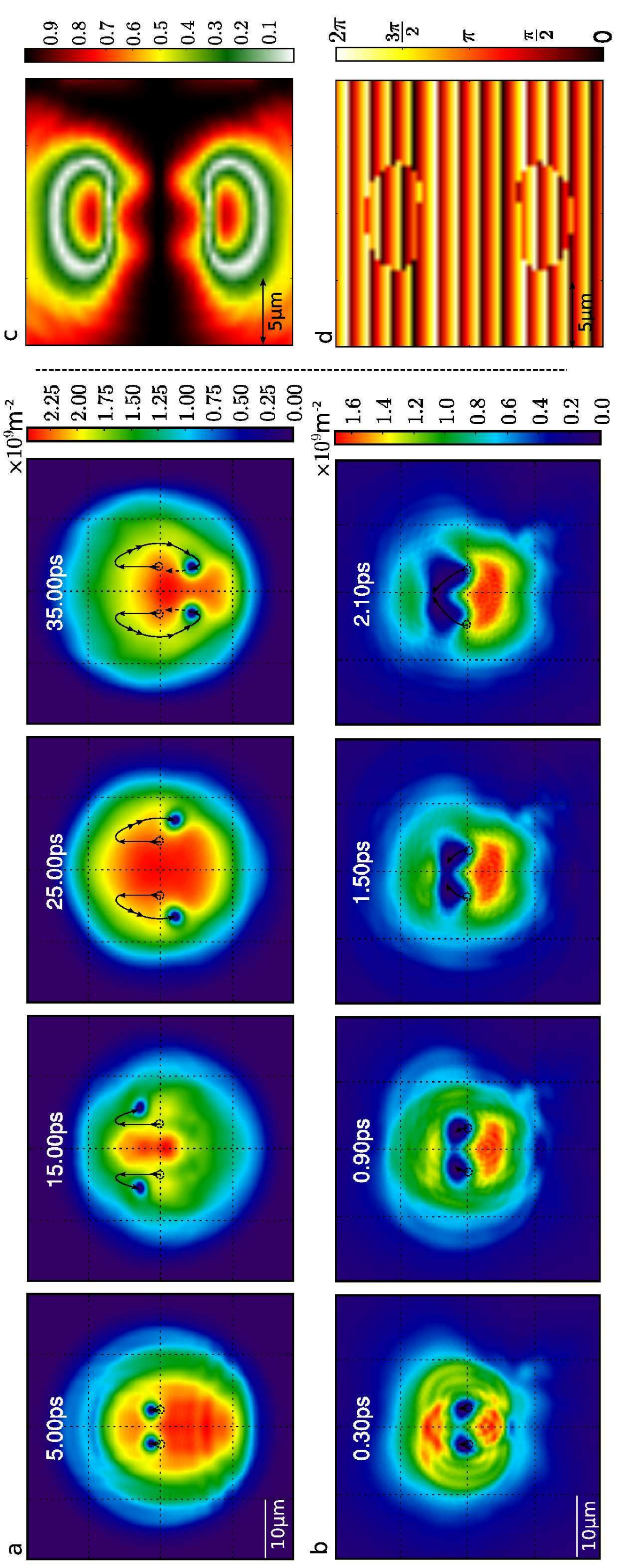
a
b
c
d
5.00ps
15.00ps
25.00ps
35.00ps
0.30ps
0.90ps
1.50ps
2.10ps
10μm
5μm
×10⁹m⁻²
2.25 2.00 1.75 1.50 1.25 1.00 0.75 0.50 0.25 0.00
1.6 1.4 1.2 1.0 0.8 0.6 0.4 0.2 0.0
0.9 0.8 0.7 0.6 0.5 0.4 0.3 0.2 0.1
2π 3π/2 π π/2 0

Figure 4:

## Single vortex-antivortex pair in an exciton-polariton condensate Supplementary Information

Georgios Roumpos[1], Michael D. Fraser[2,3], Andreas Löffler[4], Sven Höfling[4], Alfred Forchel[4] and Yoshihisa Yamamoto[1,2]

1 *Edward L. Ginzton Laboratory, Stanford University, Stanford, California 94305-4085, USA*

2 *National Institute of Informatics, 2-1-2 Hitotsubashi, Chiyoda-ku, Tokyo 101-8430, Japan*

3 *Institute for Nano Quantum Information Electronics, University of Tokyo, 4-6-1 Komaba, Meguro-ku, Tokyo 153-8505, Japan*

4 *Technische Physik, University of Würzburg Wilhelm-Conrad-Röntgen-Research Center for Complex Material Systems, Am Hubland, D-97074 Würzburg, Germany*

# 1 Supplementary Information - Experimental details

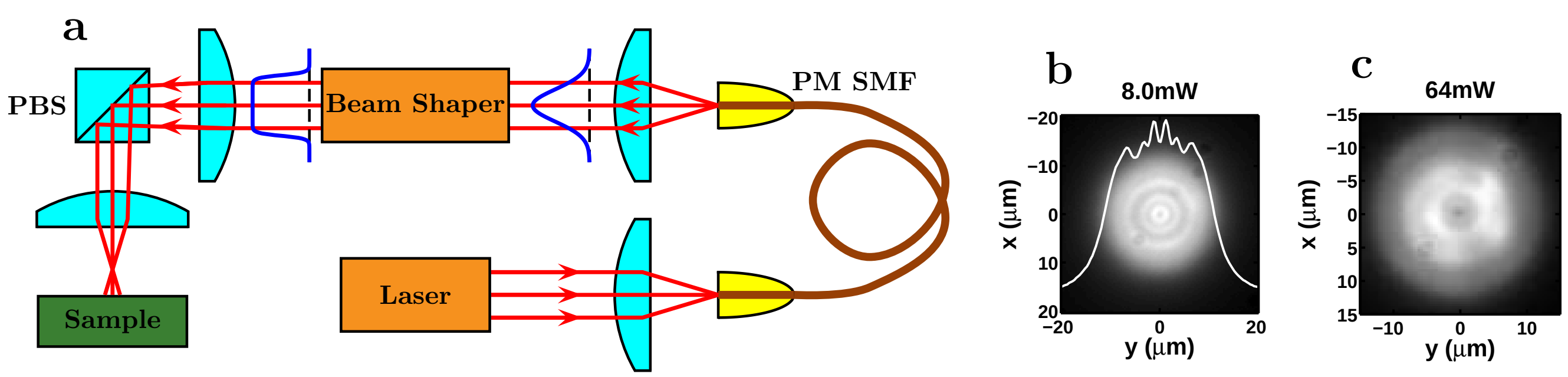

Figure 5: **Laser pumping setup.** **a**, The pumping scheme. The laser is first coupled into a polarization-maintaining single-mode fibre (PM SMF), and a collimated gaussian beam is created. We then use a commercial beam shaper to transform the beam into a collimated beam with flat-top profile. An extra lens is used in order to move the focusing point of the beam above the sample, so that a large homogeneous spot is created on it (see also Methods Summary). **b**, The pumping spot below the condensation threshold and its cross section along the horizontal axis. **c**, Real space image of the condensate above threshold.

The laser spot is created through the setup of Fig. 5a. A flat-top spot is formed (Fig. 5b) with small loss of laser power. Due to diffraction, the spot cannot be perfectly flat, but instead consists of closely-spaced bright and dark rings. For very low pumping power, the polariton diffusion length is long, and thus the rings are not visible in polariton luminescence. Above the condensation threshold, the condensate takes a doughnut-like shape (Fig. 5c), which can be reproduced by our numerical simulations (see Fig. 14).

To investigate the condensation characteristics, we perform a standard momentum- and energy-resolved measurement by forming the momentum-space image on the plane of our spectrometer slit and selecting the central stripe. In Fig. 6a, we plot the luminescence intensity measured around 0° collection angle versus the laser pump power. The spectrometer is used to discard the reflected laser light. There is a threshold pump power at $20mW$, above which the signal intensity increases a nonlinearly, similar to a lasing transition. When we record the luminescence spectrum, however (Fig. 6b), a second threshold is visible. We interpret the first threshold with polariton condensation, and the second one with photon lasing.

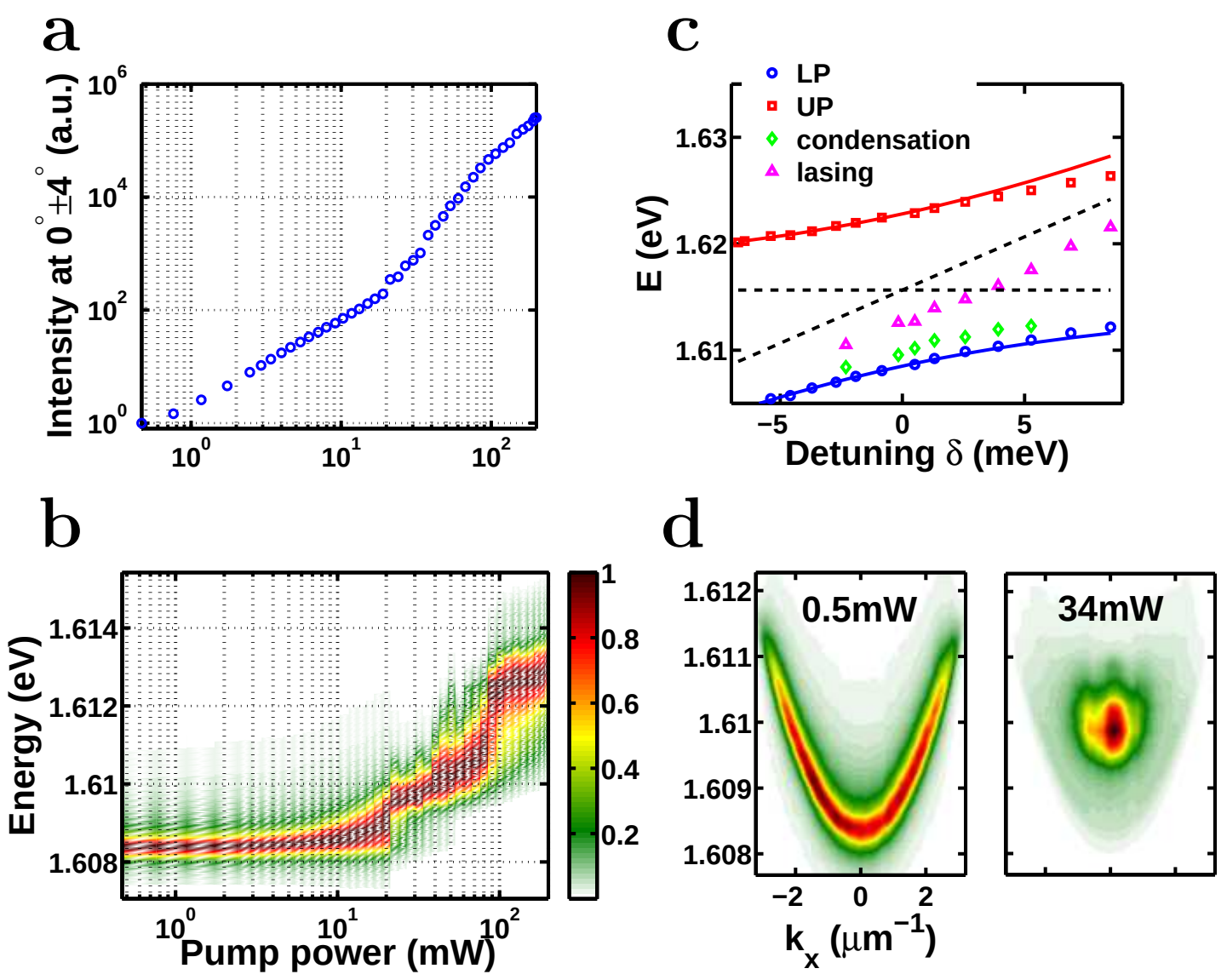

Figure 6: **Polariton condensation versus photon lasing. a**, Signal intensity measured by the spectrometer at $0 \pm 4°$ as a function of pumping power. **b**, Normalized spectrum at $0\pm4°$ as a function of pumping power. Two thresholds are visible. **c**, LP (blue circles) and UP (red squares) energies as a function of photon-exciton detuning $\delta$. Continuous lines are fits with Rabi splitting $2\hbar\Omega_R = 14meV$. The flat and tilted dashed lines are the energies of the exciton and microcavity photon respectively. The LP condensation threshold energy (green diamonds) follows the LP energies, while the photon lasing energy (magenda triangles) follows the photon energies. **d**, Dispersion relation images below and above the condensation threshold.

This interpretation is supported by Fig. 6c, which summarizes a set of measurements performed for varying photon-exciton detuning $\delta$. The wedged sample structure allows us to change the detuning by simply moving to a different position. We can observe the anticrossing between the upper polariton (UP) and lower polariton (LP) resonances. Furthermore, when we plot the energy of the LP condensation threshold (at $20mW$ in Fig. 6b, where $\delta \sim 0meV$), we see that it follows the LP resonance as shown by the green diamonds. On the other hand, the photon lasing threshold energy (at $80mW$ in Fig. 6b) follows the photon resonance as shown by the red triangles. Increasing $\delta$, so that LP's become more exciton-like, the LP condensation window shrinks, and eventually only photon lasing is possible. This data is in accordance to measurements in different samples[32,33,34]. In Fig. 6d, we plot the dispersion relations measured below and above the LP condensation threshold. A broad LP distribution is followed by a distribution peaked around zero momentum. The blue-shift is due to the repulsive interraction energy and to the decrease of the Rabi splitting because of the large quantum well exciton occupation density.

Following Ref. 5, we can estimate the healing length $\lambda_c$ from the measured blue-shift. We consider a Hartree interaction energy $gn_0 = 1meV$, so that

$$\lambda_c = \frac{\hbar}{\sqrt{2m^*_{LP} g n_0}} = 0.87\mu m. \quad (1)$$

$m^*_{LP} = 5 \times 10^{-5} m_e$ is the LP effective mass.

To make sure that the observed $\pi-$phase shift areas are not an experimental artifact, we perform the same Michelson interferometer measurement using a Gaussian pumping spot by removing the beam shaper from the setup of Fig. 5b. We found that LP condensation in this case cannot be observed with our modelocked laser, so we switch to a Ti:Sapph ring laser with long coherence time. For both prism orientations, no phase defect is observed in the interference pattern directly seen on the camera (Fig. 7a,c), or in the phase maps (Fig. 7b,d). According to the physical mechanism for creation of vortex-antivortex pairs discussed in the text, the population dip in the condensate formed because of zeros in pumping intensity is essential. This is absent when the gaussian pumping spot is used.

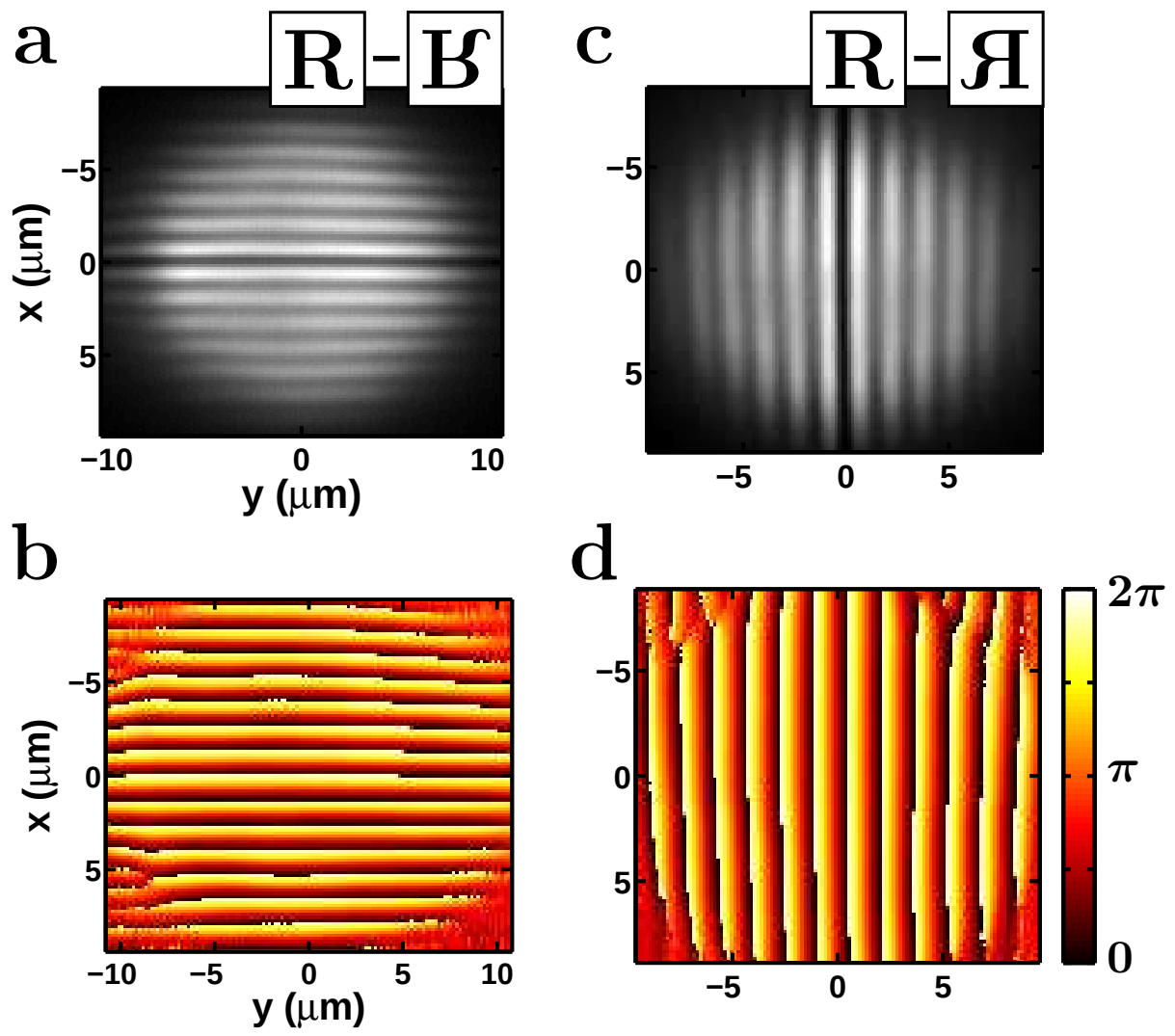

Figure 7: **Interference images with Gaussian pumping spot.** **a**, Interference fringes observed on the camera with one prism orientation. **b**, Phase map corresponding to **a**. **c**, Interference fringes observed on the camera with the orthogonal prism orientation. **d**, Phase map corresponding to **c**.

When we create a flat spot with a radius smaller by $\sim 30\%$ than the original flat spot of Fig. 5a, again no phase defects are observed. This is shown in Fig. 8, which features the measured interference patterns (Fig. 8a,c) and phase maps (Fig. 8b,d). We conclude that this condensate size is too small to support a vortex-antivortex pair.

We also rotated the original pumping spot by $90°$ using a Dove prism just before the polarizing beam splitter (PBS) in Fig. 5b. The vortex-antivortex pair is also rotated by $90°$, as is clear from Fig. 9. The two panels can be directly compared with Fig. 3(e,g) of the main text.

Finally, the observed phase defects do not depend on the observed linear polarization. We confirmed this by using a non-polarizing beam splitter instead of the PBS of Fig. 1a, along with a linear polarizer in front of the camera.

To create a map of the disorder potential, we pump with low laser power at a large spot and form the real space image of the polariton luminescence on the plane of the spectrometer slit. We can then measure the luminescence spectrum along a line on the sample (Fig 10a). By moving the real space image to the direction normal to the slit, we can measure the spectrum of a two-dimensional grid of points. Our resolution is $1\mu m$. We then fit the long-wavelength part of the spectrum with one half of a Lorentzian and extract the local energy for every point (Fig 10b). We find a striped pattern for the disorder potential (Fig 10c), while the local energy follows a Gaussian distribution with $\sigma = 71\mu eV$ (Fig 10d).

Using a prism in our Michelson interferometer (M2 in Fig. 1a of the main text) instead of a retroreflector has the advantage that a single vortex should always be observable, even if it is mobile. The idea is illustrated in Fig. 11a and is based on the fact that the sense of rotation changes when an image is reflected. Consider the case that the original image includes one vortex. The reflected image will include one antivortex. Because the interferometer measures the phase difference between the two images, the final image will feature two vortices. Fig. 11b shows the simulated time-integrated interference pattern when a single vortex is moving randomly inside the red circle. Fig. 11c shows the same interference pattern for the case of a mobile antivortex, while in Fig. 11d we consider a 50-50 probability for either vortex or antivortex. None of these interference patterns is observed in our data, which suggests that there are no free vortices in the condensate, but only bound vortex-antivortex pairs.

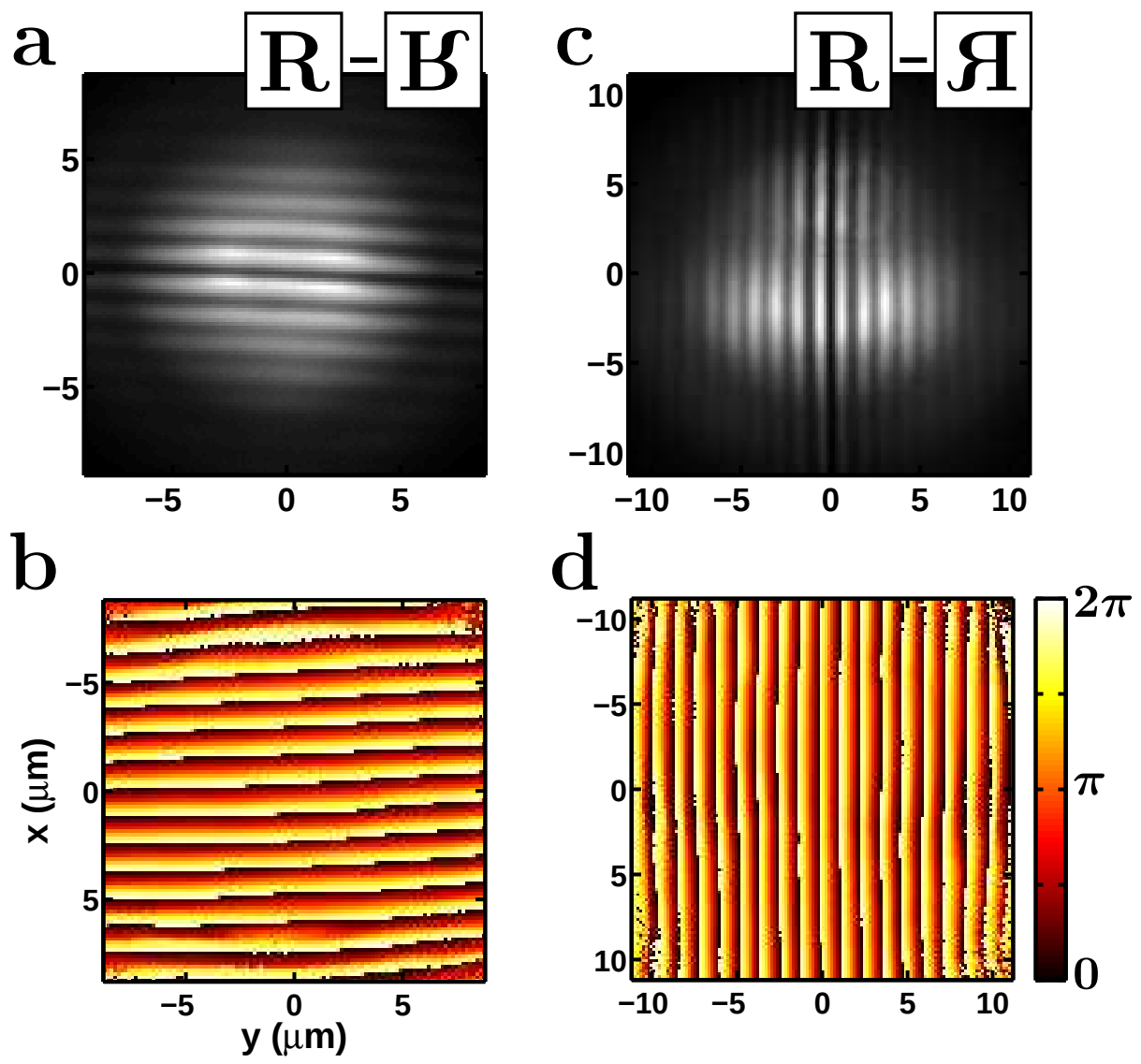

Figure 8: **Interference images with a small flat pumping spot.** **a**, Interference fringes observed on the camera with one prism orientation. **b**, Phase map corresponding to **a**. **c**, Interference fringes observed on the camera with the orthogonal prism orientation. **d**, Phase map corresponding to **c**.

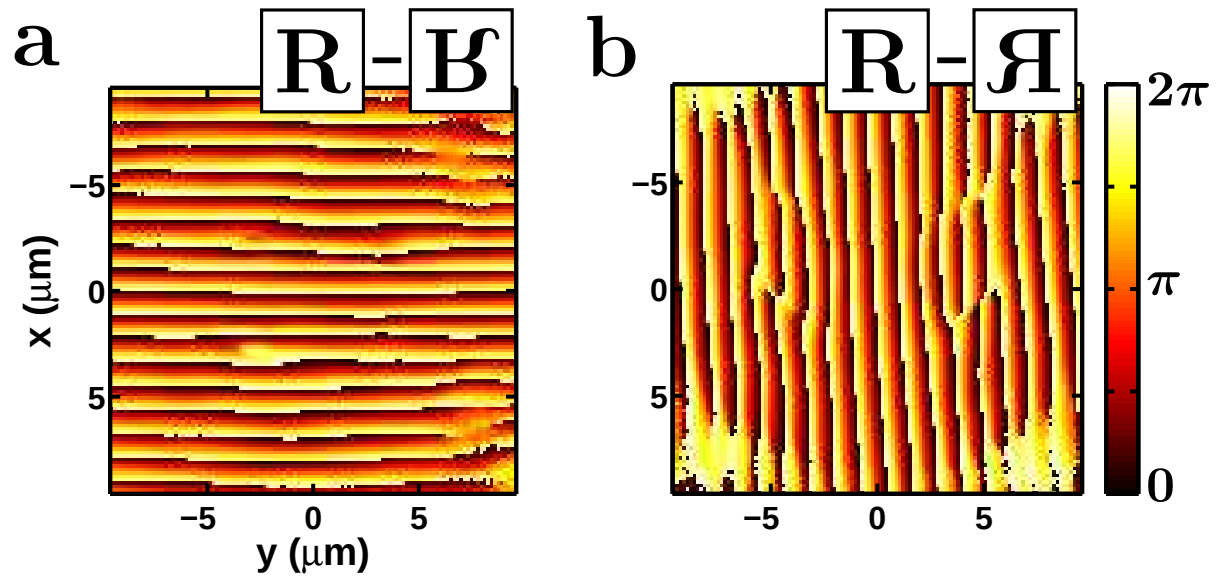

Figure 9: **Interference images with a pumping spot rotated by** $90°$**.** **a**, Phase map with one prism orientation. **b**, Phase map with the orthogonal prism orientation.

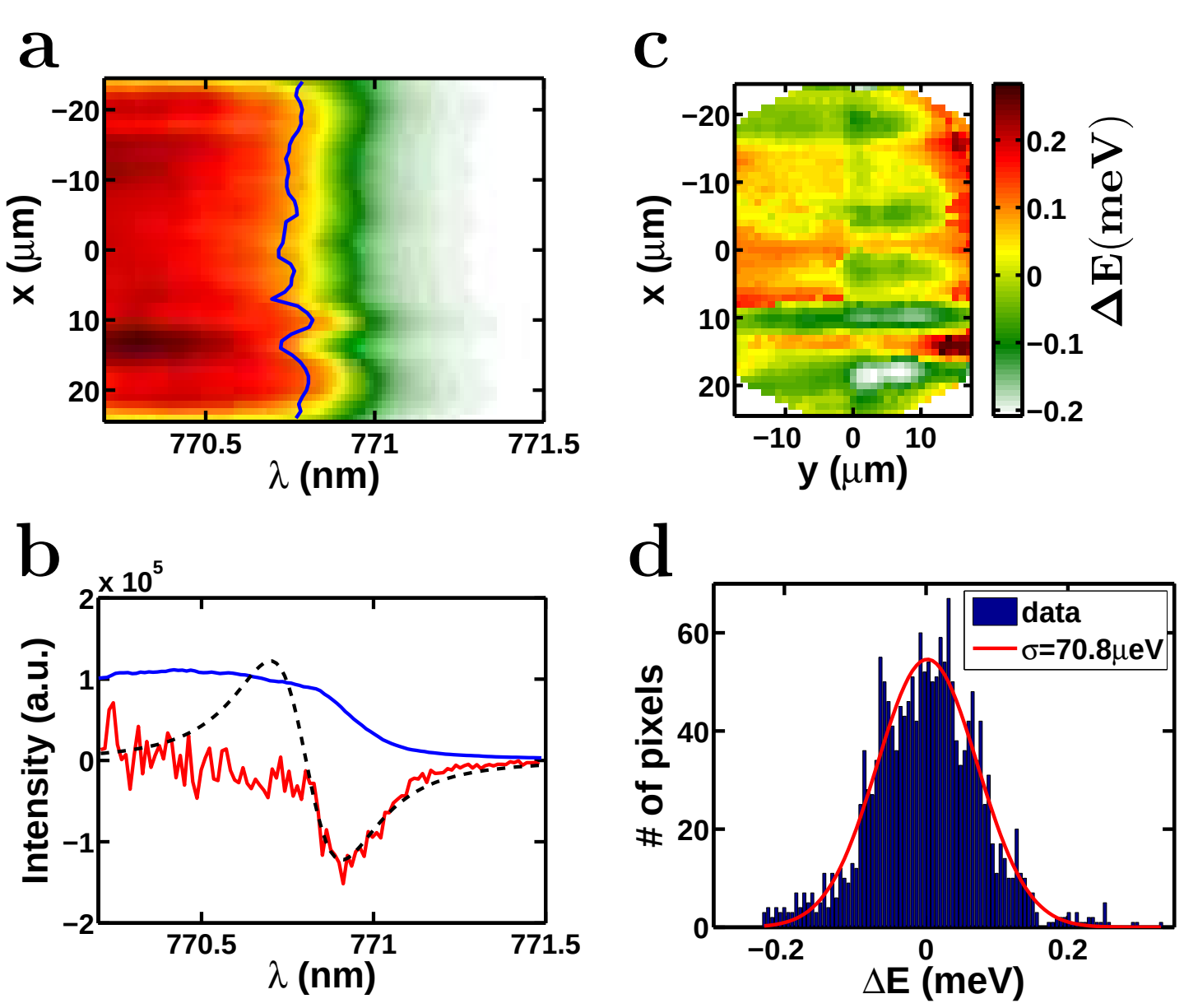

Figure 10: **Sample disorder potential.** **a**, Example of spatially-resolved spectra. The blue line is the local wavelength selected by our fitting method. **b**, The fitting method. Blue: spectrum selected from **a** along $x = 0\mu m$. Red: numerical derivative. Black dashed line: fitting of the long-wavelength part of the numerical derivative with the derivative of a Lorentzian. **c**, Map of the disorder potential. White areas at the corners have low signal to noise ratio. **d**, Histogram of the disorder potential shown in **c**, and fit with a Gaussian with $\sigma = 70.8\mu eV$.

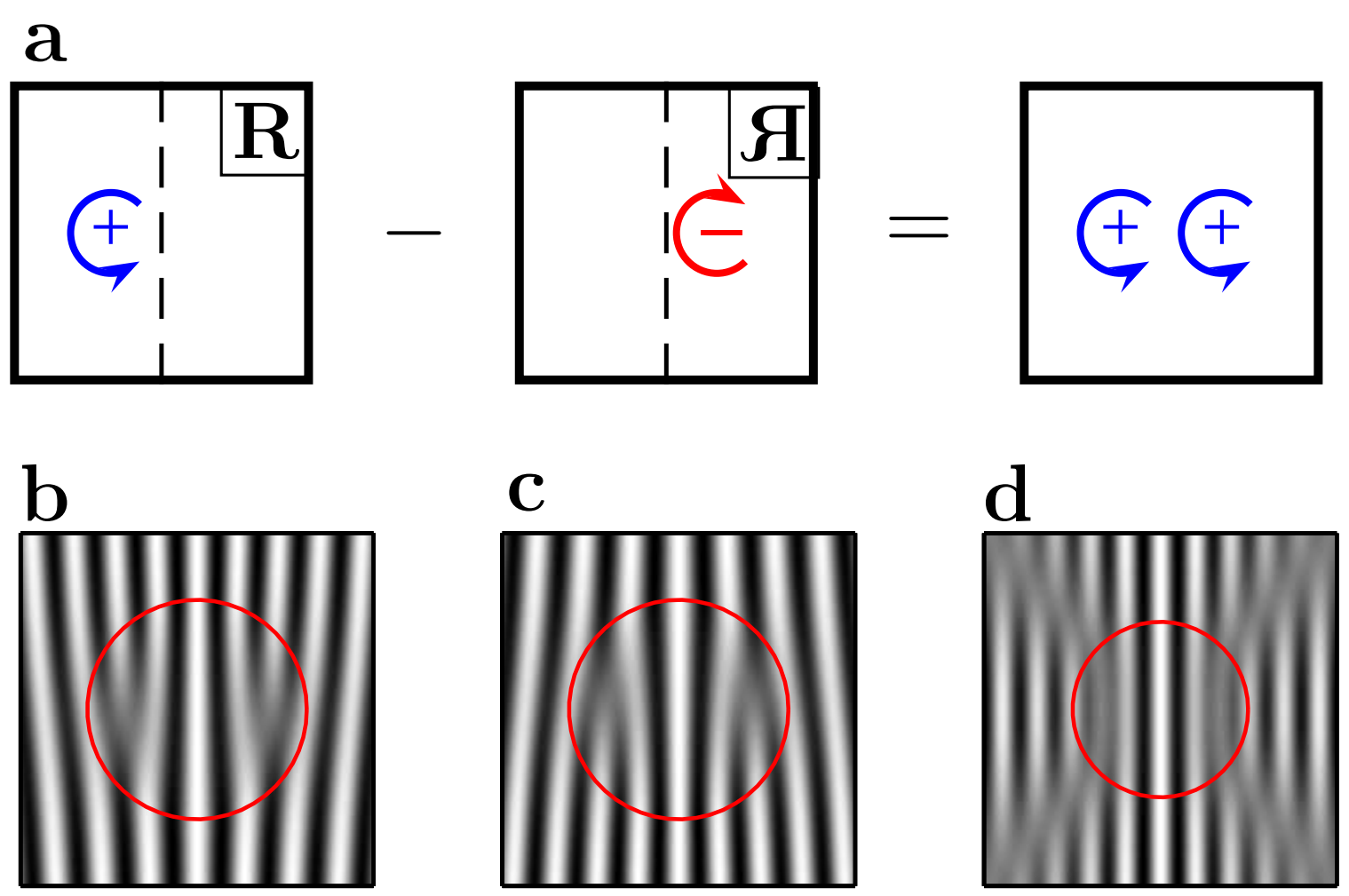

Figure 11: **Interference images for a single mobile vortex.** **a**, Schematic of the Michelson interferometer measurement for the case that a single vortex is present in the initial image (see text). **b-c**, Simulated time-integrated interference patterns for the case that a single vortex (b) or a single antivortex (c) are moving randomly inside the red circle. **d** The same as b and c, but now there is a 50-50 probability for vortex or antivortex.

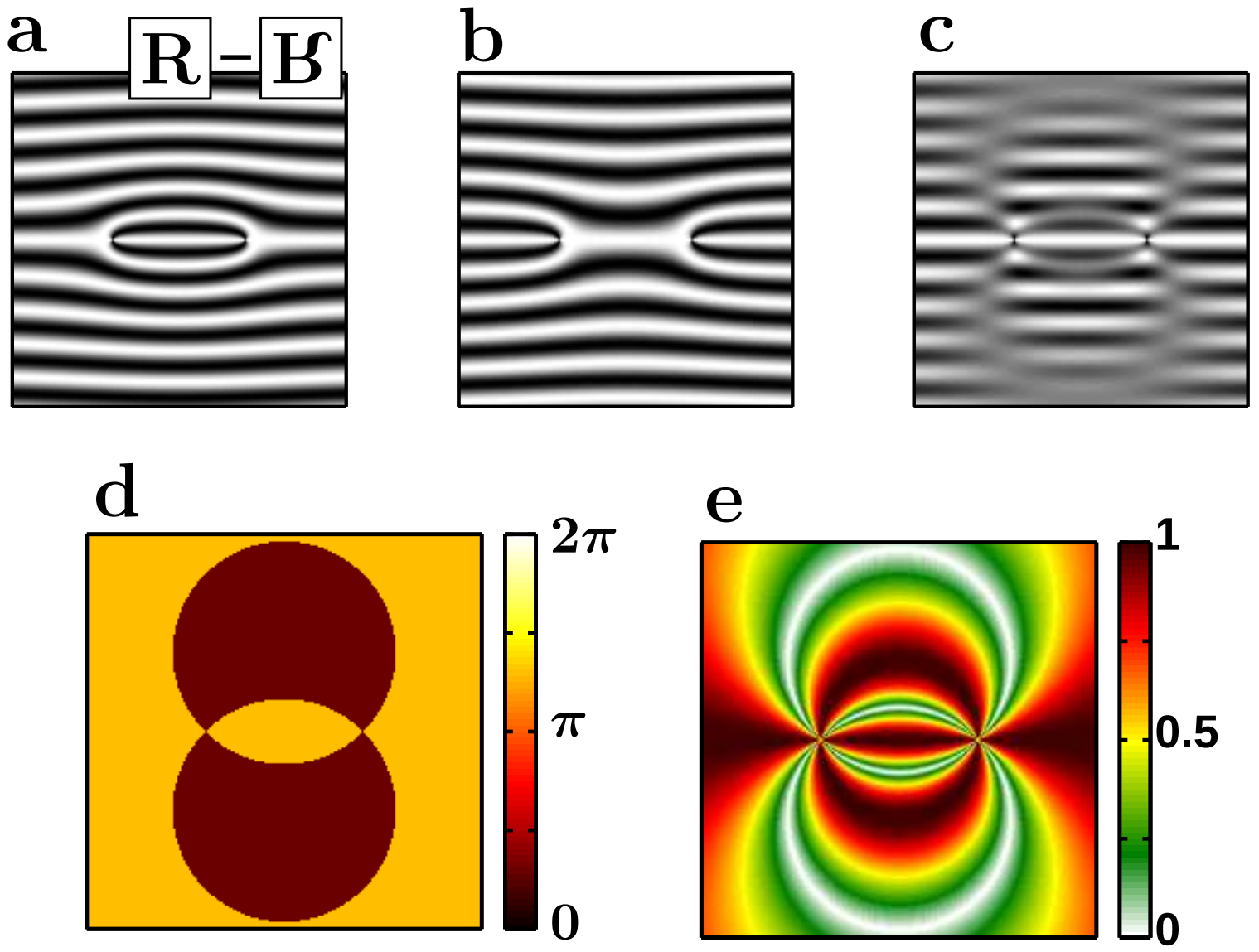

Figure 12: **Explanation of the interference pattern for a single pair with random polarization (simulation results).** **a**, Interference fringes observed in the camera for one vortex-antivortex pair polarization and for vertical orientation of the prism in the Michelson interferometer. It features two trident-like patterns. **b**, Interference fringes for the opposite pair polarization. The trident-like patterns are now reversed. **c**, Interference fringes for 50-50 statistical mixture of the two pair polarizations. **d**, Phase map and **e**, visibility map corresponding to c.

Fig. 12 analyzes the characteristic pi-phase-shift interference pattern observed for a single vortex-antivortex pair with random polarization. Fig. 12a shows the interference fringes we expect to observe on the camera when there is a single pair at a fixed position along the horizontal axis and the prism in the Michelson interferometer is oriented vertically. Two trident patterns appear. When the pair polarization is reversed (Fig. 12b), the trident patterns are also reversed. Fig. 12c results from 50-50 statistical mixture between the two pair polarizations. There are two areas where the phase of the fringes is shifted by pi compared to outside, and around these areas the fringes are blurred. This is because in case a the fringes are bent one way, whareas in case b the fringes are bent the other way. Figs. 12d,e show the phase and visibility maps corresponding to Fig. 12c, where the pi-phase-shift areas are clearly seen. Obviously, the shape of these areas depends on the position of the pair, and should be different when the pair can move inside the condensate without changing its orientation, as will be discussed later. We expect that this motion takes place in our experiments.

Fig. 13 explains the qualitative difference between the two prism orientations when the vortex-antivortex pair lies along the horizontal axis. In one case, the vortex and antivortex in the reflected image annihilate those in the original image, whereas in the other case the final image features a double vortex and a double antivortex. This is the reason why Fig. 3g of the main

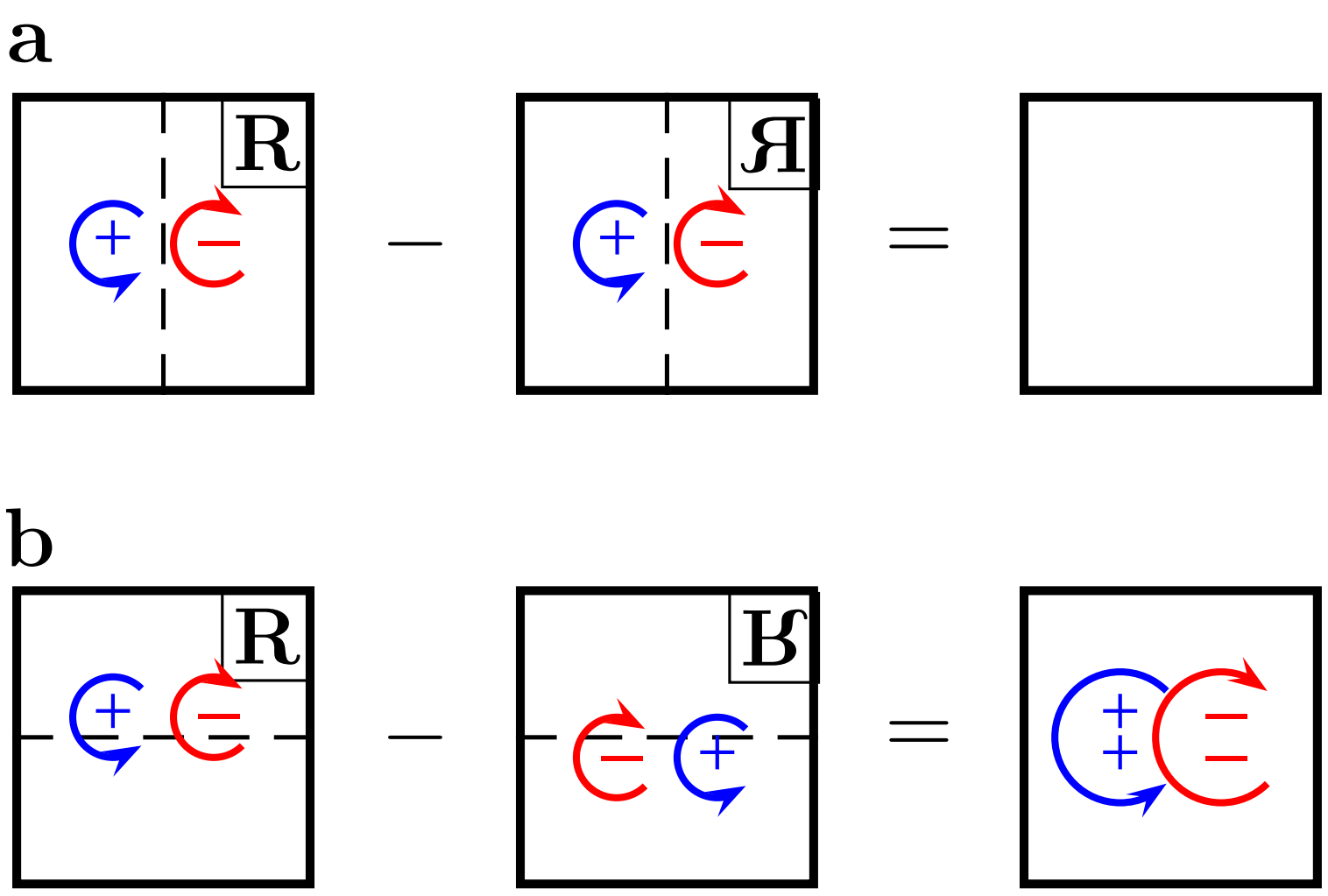

Figure 13: **Horizontal versus vertical prism orientation.** **a**, When the vortex-antivortex pair lies along the horizontal axis and the original image interferes with its reflection along the vertical axis, then the vortex and antivortex in the reflected image annihilate those in the original image. So the final image does not show any phase defects, corresponding to Fig. 3g of the main text. **b**, When the prism creates the reflection of the original image along the horizontal axis, then the final image shows a double vortex and a double antivortex, corresponding to Fig. 3e of the main text. This is a qualitative change with respect to a.

text does not show any phase defect in contrast to Fig. 3e.

## 2 Vortex nucleation and pair dynamics theory

In the main text we have described the observation of a single dynamic vortex-antivortex pair in a dissipative polariton condensate. Though the time-integrated experiment results in clear interference fringes indicating the presence of a single vortex-antivortex pair, we have implemented a theoretical model and performed numerical simulations to understand the dynamics and nucleation mechanism and to gain further insights. The model is found to reproduce the essential part of these observations.

### 2.1 Open-dissipative Gross-Pitaevskii equation

While atomic condensates display dynamic features on timescales (order of $ms$) much shorter than the condensate lifetime (order of $s$), the shortness of the polariton lifetime $\tau_{pol} \sim 3ps$ in addition to condensate-reservoir interactions are expected to significantly influence the vortex pair lifetime and dynamics in this system. Thus in order to understand the dynamics of a vortex-antivortex pair in a polariton condensate, the usual form of the Gross-Pitaevskii equation[25] is insufficient. In a conservative confined condensate, the vortex pair motion is governed by the effective mass $m_{LP}$ and interaction parameters $g_C n_C$ (determining the healing length $\xi = \hbar/\sqrt{2m_{LP}g_C n_C} \sim 0.9\mu m$ and speed of sound $v_s = \sqrt{g_C n_C/m_{LP}} \approx 2\mu m/ps$) and the condensate size / trapping potential. In contrast, the use of a two-level dissipative Gross-Pitaevskii equation (GPE)[29] to model the polariton condensate dynamics allows the study of the influence of the further parameters such as condensate polariton loss rate $\gamma_C$ and stimulated scattering rates $R(n_R(\mathbf{r}, t))$.

The open-dissipative Gross-Pitaevskii model consists of the following coupled equations describing the time evolution of the condensate order parameter $\psi(\mathbf{r},t)$ and reservoir polariton density $n_R(\mathbf{r},t)$

$$i\hbar\frac{\partial\psi(\mathbf{r},t)}{\partial t} = \left(-\frac{\hbar^2\nabla^2}{2m_{LP}} + V_{ext}(\mathbf{r}) - \frac{i\hbar}{2}\left[\gamma_C - R(n_R(\mathbf{r},t))\right] + g_C|\psi(\mathbf{r},t)|^2 + g_R n_R(\mathbf{r},t)\right)\psi(\mathbf{r},t) \tag{2}$$

$$\frac{\partial n_R(\mathbf{r},t)}{\partial t} = P_{las}(\mathbf{r},t) - \gamma_R n_R(\mathbf{r},t) - R(n_R(\mathbf{r},t))|\psi(\mathbf{r},t)|^2 \tag{3}$$

The reservoir density $n_R(\mathbf{r},t)$ is further controlled by laser pumping gain $P_{las}(\mathbf{r})$ and reservoir loss $\gamma_R$ parameters. Interaction between condensate and reservoir polaritons is assigned a coupling constant $g_R = 2g_C$, where the condensate coupling constant is $g_C \simeq 6\times10^{-3} meV\mu m^2$. This model has previously been used successfully to describe the excitation spectrum[29], and to study experimental results for a pinned vortex in a polariton condensate[16].

## 2.2 Vortex-antivortex pair dynamics

We have analyzed the dynamics and stability of a vortex-antivortex pair in a polariton condensate in detail[30] and demonstrated significant deviations from the usual vortex pair motion[35] expected in the conservative atomic condensate. We found that the vortex pair will either recombine within the condensate or separate and dissipate from the condensate boundary. The choice between these two types of trajectory (in a largely homogeneous but confined condensate) is a competition between the force[36,35] due to the radially directed superfluid flow of the unconfined repulsively interacting condensate, and the drag forces due to the interaction with non-condensate (reservoir and thermal polaritons)[37]. The cross-over between these regimes essentially depends only on the magnitude and profile of $n_R(\mathbf{r})$.

The simulation pumping profile used is a top-hat experimentally measured profile (see Fig. 5b) and the condensate polariton lifetime is $\tau_r = 3ps$. The scattering rate $R(n_R) = R_{sc}n_R(\mathbf{r})$ is assigned a linear dependency and a measurement of the threshold pumping power $P_{th}$ permits an estimate of the scattering rate via $R_{sc} = \gamma_C\gamma_R/P_{th}$, where we assume $\gamma_R$ is comparable to $\gamma_C$, necessary to study this experimental parameter space. Thus, with these parameters fixed to correspond to experiments, a variation in the normalized pumping power $\bar{P} = P_l/P_{th}$ notably alters the relative fraction of reservoir $n_R$ to condensate $n_C$ particles according to $\frac{n_R}{n_C} = \frac{\gamma_C}{\gamma_R}\frac{1}{\bar{P}-1}$, and thus the pumping power should also control the choice between the two possible vortex pair trajectories outlined in reference 30.

Additionally specific to these experiments, as the healing length of the condensate is not significantly smaller than the condensate size, the condensate boundaries are expected to also affect the vortex pair motion. Depending on the vortex pair initial energy and the local condensate environment, the process of vortex pair either recombining within the condensate or splitting and leaving the condensate will happen on the order of the condensate polariton lifetime due to the small condensate size. The next section aims to establish the position of these experiments in the possible parameter space and with this determine vortex pair motions and attempt to reproduce experimental steady-state measurements.

## 2.3 Steady-state condensate profiles and time-integrated measurements

In the absence of any external confinement, the steady-state condensate profile $n_C(\mathbf{r})$ is largely determined by stimulated scattering $R(n_R(\mathbf{r}))$ in turn driven by the laser profile $P_l(\mathbf{r})$, and by interactions between the condensate and reservoir polariton populations. When $n_R$ is small, $n_C(\mathbf{r})$ will closely mimic the $n_R(\mathbf{r})$ and $P_l(\mathbf{r})$ profiles. However, when the magnitude of $n_R(\mathbf{r})$ is comparable to that of $n_C(\mathbf{r})$, the profile of $n_C(\mathbf{r})$ will be strongly influenced by the need to minimize the interaction term $g_R n_R(\mathbf{r},t)\psi(\mathbf{r},t)$.

To clarify this, we consider the experimental pump spot (Fig. 14a) and simulate the corresponding reservoir (Fig. 14b) and condensate (Fig. 14c) distributions, which show clearly the inverse correspondence between population inhomogeneities (of diffusion length range) in the condensate and reservoir polariton profiles. In this simulation, we have deliberately chosen pumping parameters which result in large $n_R(\mathbf{r})$ to reproduce the experimental steady-state condensate profile (compare Fig. 14c and Fig. 5c). This notably includes the feature of a spatial dip at the condensate origin. Though experimentally it is difficult to know exactly where in the parameter space the condensate is situated (due largely to a lack of exact knowledge of $R(n_R(\mathbf{r}))$), the inhomogeneities in the experimentally measured steady-state condensate profile can only be reproduced by permitting a large reservoir density. The fact that measurements with pumping powers up to $\bar{P} \approx 10$ have been performed, all showing similar strong spatial modulations resulting from pump inhomogeneities confirms our choice of dissipative parameters $\gamma_R$ and $R_{sc}$ and that even pumping far above threshold, the reservoir is still expected to play a significant role. This then suggests that the vortex dynamics will be strongly modified by drag forces and thus pair recombination is preferred over pair splitting. It is found that this feature of our experiment is crucial in determining the formation and subsequent life cycle of the vortex-antivortex pair as noted in the main text.

Though the experimental pump profile (Fig. 14d) is roughly symmetric, existing asymmetries are enhanced through the presence of the large repulsive interaction between reservoir and condensate modes, resulting in an asymmetric condensate profile in the numerical simulation (Fig. 14e). This result biases the formation of vortex pairs preferentially dipole aligned with one Cartesian axis, in this case the horizontal axis. Furthermore, the conclusion that an average of only one vortex pair is present is to be expected based on our experimental results and from the observation that the minimum vortex pair size ($2d_v \sim 4\mu m$) is comparable to the size of the central dip in the condensate, believed responsible for vortex pair formation.

It is also desirable to know in more detail how the dynamics and life cycle of a single vortex-antivortex pair are reflected in the time-integrated measurements, and specifically why the defects in the fringes are not washed out by vortex motion. We assume that the vortex pair (imprinted in the calculation directly via a phase factor $e^{il\theta}$ where $\theta = tan^{-1}(\frac{y}{x \mp d_v/2})$ and $l = \pm 1$) is formed along the $x$-axis with core-core separation $d_v$ and is subsequently free to evolve in time. Essentially, given to the mirror-symmetry of the problem across the $y$-axis, the interference fringes will be observed due to the correlated motion of the vortex and antivortex. The system topology restricts the pair motion to identical velocities in the same $y$-direction and opposite $x$-direction, with no pair rotation. The dislocations in the interference measurement and visibility minima patterns are easily reproduced despite vortex pair motion given that (a) the mirror symmetry of the vortex pair about its midpoint and (b) that the condensate is occupied by a vortex pair most of the time. The shape of these $\pi$-phase shifted regions in the interference and fringe visibility experiments and simulations have a strong dependence on the vortex pair dynamics, specifically a correspondence between the area of the $\pi$-phase shifted region and area mapped by the vortex pair trajectory. Small $\pi$-phase shifted areas (as in experiments) are only reproduced with recombining vortex pairs, while separating vortex pairs generally give distinct and significantly larger areas.

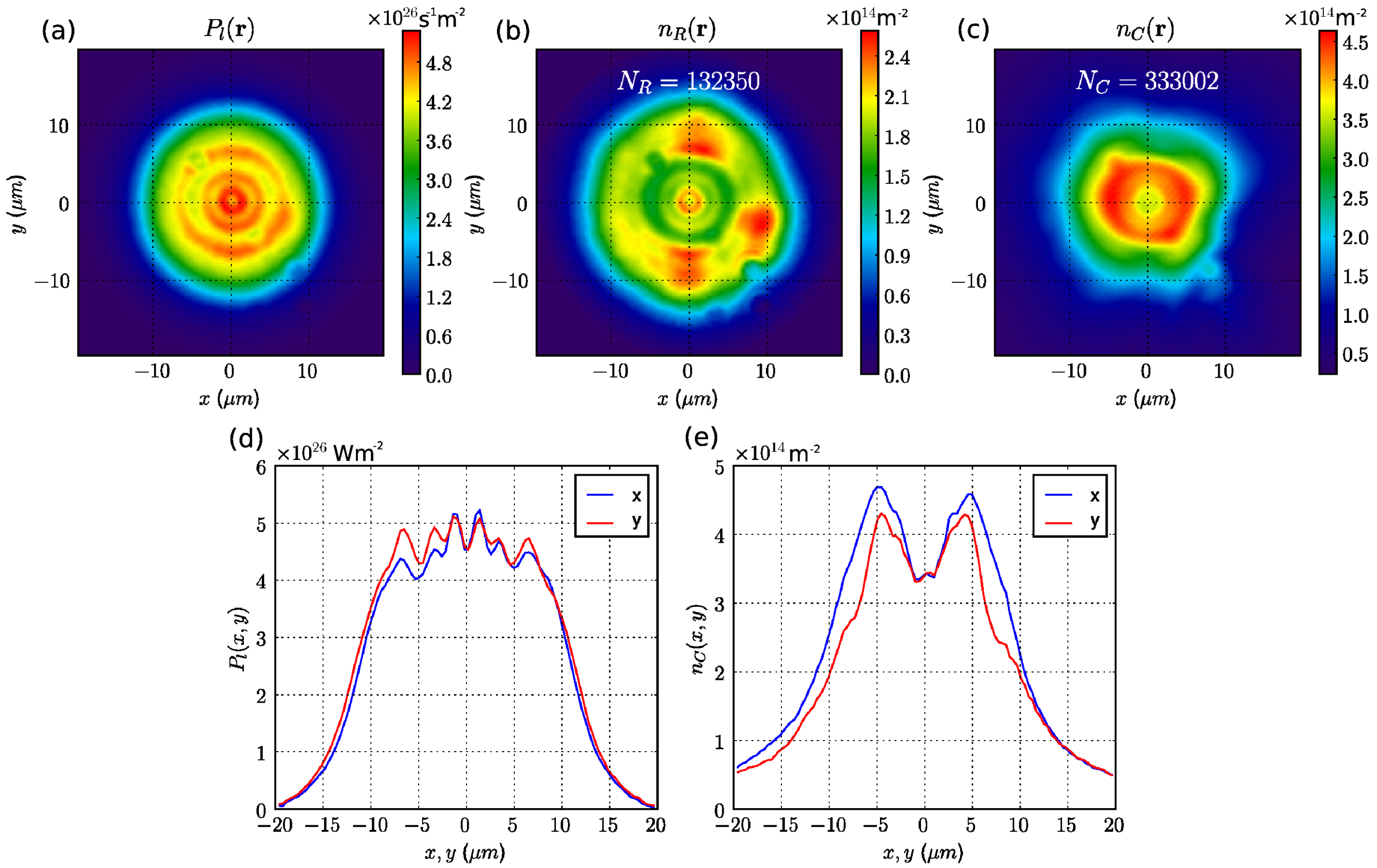

Figure 14: The experimental pump profile (a), when used in the numerical simulations produces steady-state reservoir (b) and condensate (c) polariton profiles. The number of polaritons in the reservoir $N_R$ and condensate $N_C$ are similar inducing the inverse correspondance in these two profiles by the corresponding strong repulsive interactions. Though the difference between the $x$ and $y$ axis of the laser profile (d) is small, repulsive interactions produce a strongly asymmetric steady-state condensate profile (e).

Thus, two pieces of evidence, namely (a) strong modulation of condensate spatial profile by an inhomogeneous pump and (b) the requirement that the vortex pair recombines rather than splits to successfully explain and simulate the experimental results both imply a similar parameter space position, namely condensate and vortex dynamics are strongly influenced by interaction with reservoir populations.

## 2.4 Probability of spontaneous vortex formation

In two-dimensions, the spontaneous formation of vortex pairs is usually the result of interaction of the condensate population with thermal phase fluctuations[25]. Knowing the energy $E_v$ of a particular vortex pair configuration, the probability of thermal nucleation of this vortex pair configuration can be estimated. In three dimensions, the thermal excitation of a single vortex requires an energy proportional to the length of the vortex line, and is thus highly improbable in both atomic and polariton systems. In two dimensions (2D) though, the confinement along the axis of the vortex line results in a much smaller formation energy. In the present polariton condensate, the thermal energy is still insufficient to generate a single vortex which we can estimate from equation

$$E_v \sim \frac{m_{LP} n_C \kappa^2}{4\pi} \ln(\frac{L}{\xi}) \sim 10eV \gg k_B T \tag{4}$$

for the single on-axis vortex in an inhomogeneous condensate[38]. In this equation, $L$ is the condensate radius and $\kappa = h/m_{LP}$. When dissipative effects are taken into account (through the dissipative GPE), the vortex energy is found to be lower than this estimate and non-linear in $n_C$ due to the reservoir presence, but is still of order $eV$. The energy of a 2D vortex-antivortex pair though, depends on its separation and thus can have a smaller energy for small separations. Equation

$$E_p = \frac{m_{LP} n_C \kappa^2}{2\pi} \ln\left(\frac{d_v}{r_0}\right) \tag{5}$$

gives the energy for a uniform condensate. Corrections to this equation for a finite size condensate are only important for vortex pairs within distances $\xi$ of the boundary. Unless the vortex pair is given enough energy for the vortex cores to fully separate (core-core separation $d_v \gtrsim 2\xi$ it will immediately recombine rapidly and will not be observed in the current type of experiments. Thus, unless the system contained a great many vortex pairs simultaneously, the rapid formation/annihilation of small vortex pairs will have a negligible effect on measurements and dynamics.

The coefficient of the logarithm is still large however, and thermal fluctuations ($k_B T \sim 0.4meV$) of the phase cannot to be solely responsible for vortex pair formation. Instead, density fluctuations in the thermal reservoir which has population density comparable to the condensate permits an alternative formation source as $g_R n_R \sim \mu$. These fluctuations in the reservoir density result from the use of a noisy CW multi-mode laser with strong amplitude fluctuations expected at a timescale of the order of $ps$, which is similar to vortex pair lifetime. In this case, the central dip in the condensate profile discussed previously presents ideal conditions for vortex pair formation.

Phase fluctuations[25,28,39] are included in the operator of the condensate order parameter through $\hat{\psi} = \sqrt{n_C(\mathbf{r})} e^{i\hat{S}(\mathbf{r})}$, where $n_C(\mathbf{r})$ again represents the condensate density distribution at $T = 0$ and $S(\hat{\mathbf{r}})$ is the phase fluctuation operator. A phase correlation function can then be defined as $\chi(\mathbf{r} - \mathbf{r}') = \langle S(\hat{\mathbf{r}}) S(\hat{\mathbf{r}'}) \rangle$.

For a 2D finite area Bose gas, the phase fluctuation operator $\hat{S}(\mathbf{r})$ can be expressed in terms of the usual Bogoliubov coefficients $u_k(\mathbf{r})$ and $v_k(\mathbf{r})$, with

$$\hat{S}(\mathbf{r}) = \frac{1}{4\sqrt{n_C(\mathbf{r})}} \sum_k \left[ f_k^+(\mathbf{r}) \hat{a}_k + f_k^-(\mathbf{r}) \hat{a}_k^\dagger \right] \tag{6}$$

where $\hat{a}_k$ is the annihilation operator of an excitation quanta from state $k$, where the amplitudes are $f_k(\mathbf{r})^\pm = u_k(\mathbf{r}) \pm v_k(\mathbf{r})$. A full treatment for the dissipative polariton condensate should also include density fluctuations and explicit contributions from interaction with reservoir populations, but the essential statement relevant here is that $\chi(\mathbf{r} - \mathbf{r}')$ is locally maximized when $n_C(\mathbf{r})$ is minimized and the non-condensate population (thermal or reservoir) is maximized. Thus the local exchange of energy from condensate to reservoir is expected to be maximized at this position and the largest probability of phase fluctuations occurs, which are the specific conditions at the centre of the excitation spot in present experiments.

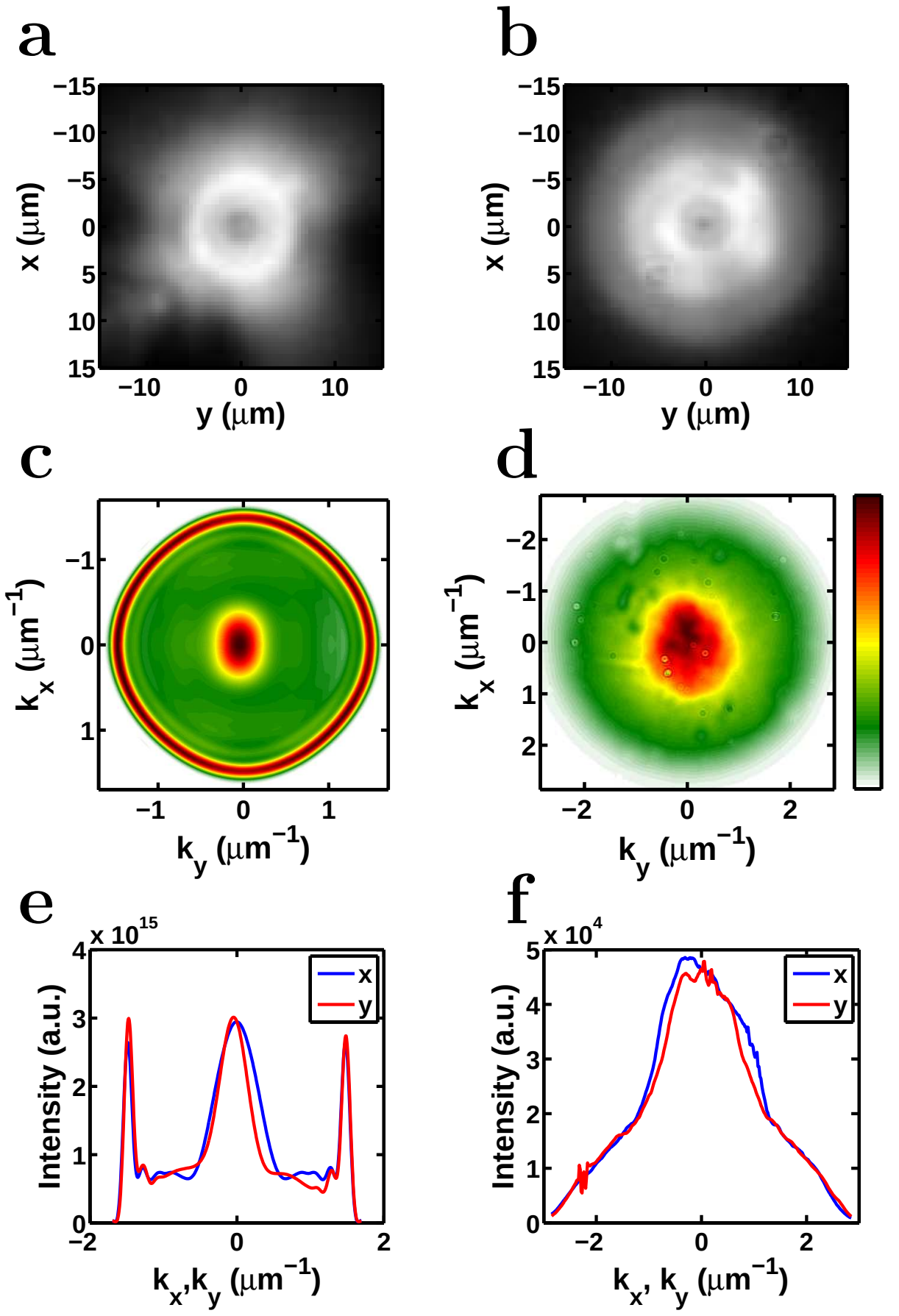

Figure 15: **Real- and momentum-space images.** **a**, Simulation result for the real space distribution of the condensate without the presence of a vortex-antivortex pair. **b**, Measured real space image above the condensation threshold (at 55 mW). **c**, Simulated momentum space image corresponding to a with an imprinted vortex-antivortex pair. The possibility that vortex and antivortex positions swap is included. **d**, Measured momentum space image, averaged over a grid of 15 points on the sample. **e**, Cross sections of the momentum distribution in c. **f**, Cross sections of the momentum distribution in d.

# 3 Real and momentum space images

For a pinned vortex-antivortex pair, one expects a real space distribution with two minima, marking the position of the vortex and antivortex. Our experimental time-integrated result (Fig. 15b) instead does not feature this pattern. This suggests that the observed pair is not pinned at any particular position. Fig 15a shows the simulation result of our dissipative model. In a superfluid, on the other hand, phase gradient induces superfluid motion. Because pairs in our experiment sit along a fixed axis, the momentum space distribution should be asymmetric. In Fig. 15c we show the calculated momentum space distribution corresponding to Fig. 15a with an imprinted vortex-antivortex pair. We again allow for the possibility that the vortex and antivortex can swap positions. The distribution is elongated perpendicular to the direction of the pair. The peak at $|k| \sim 1.5\mu m^{-1}$ is due to LP's travelling away from the condensate. In the experiment, we expect the momenta of these LP's to follow a broad distribution due to mutual scattering. In Fig. 15d we present the measurement result for the momentum space distribution above the condensation threshold (at 55mW). Because of the small sample disorder potential, the momentum space distribution is always slightly inhomogeneous. To suppress this effect, we average the measured distributions over a grid of 15 equally spaced points on the sample. The cross sections along the two axes corresponding to Fig. 15c,d are shown in Fig. 15e,f respectively. The predicted small asymmetry is indeed observed. After rotating the sample by $90°$, the asymmetry remains along the same direction, which confirms that the effect is not due to the sample disorder potential.